\documentclass{aastex}          
\usepackage{spr-astr-addons}    

\newcommand{\as}{$\arcsec$}
\newcommand{\bz}{{\it Roma}-BZCAT}
\newcommand{\eg}{{\em e.g.}}
\newcommand{\frm}{{\it Fermi}}
\newcommand{\ie}{{\em i.e.}}
\newcommand{\sw}{{\it Swift}}
\newcommand{\wse}{WISE}

\begin{document}

\title{New blazars from the cross-match of recent multi-frequency catalogs}

\shorttitle{New blazars from the cross-match of recent multi-frequency catalogs}
\shortauthors{A. Maselli et al.}

\author{A.~Maselli\altaffilmark{1}} 
\altaffiltext{1}{INAF-IASF Palermo, via U.~La~Malfa 153, I-90146 Palermo, Italy}
\email{maselli@ifc.inaf.it} 
\and
\author{F.~Massaro\altaffilmark{2,3}}
\altaffiltext{2}{Dipartimento di Fisica, Universit\`a degli Studi di Torino, via Pietro Giuria 1, I-10125 Torino, Italy}
\altaffiltext{3}{Yale Center for Astronomy and Astrophysics, Physics Department, Yale University, PO Box 208120, New Haven, CT 06520-8120, USA}
\and
\author{R.~D'Abrusco\altaffilmark{4}}
\altaffiltext{4}{Harvard-Smithsonian Astrophysical Observatory, 60 Garden Street, Cambridge, MA 02138, USA}
\and 
\author{G.~Cusumano\altaffilmark{1}} 
\altaffiltext{1}{INAF-IASF Palermo, via U.~La~Malfa 153, I-90146 Palermo, Italy}
\and 
\author{V.~La~Parola\altaffilmark{1}} 
\altaffiltext{1}{INAF-IASF Palermo, via U.~La~Malfa 153, I-90146 Palermo, Italy}
\and 
\author{A.~Segreto\altaffilmark{1}}
\altaffiltext{1}{INAF-IASF Palermo, via U.~La~Malfa 153, I-90146 Palermo, Italy}
\and
\author{G.~Tosti\altaffilmark{5,6}}
\altaffiltext{5}{Dipartimento di Fisica, Universit\`a degli Studi di Perugia, I-06123 Perugia, Italy}
\altaffiltext{6}{Istituto Nazionale di Fisica Nucleare, Sezione di Perugia, I-06123 Perugia, Italy}

\begin{abstract}
Blazars are radio-loud active galactic nuclei well known for their non
thermal emission spanning a wide range of frequencies.
The \bz\ is, to date, the most comprehensive list of these sources.
We performed the cross-match of several catalogs obtained from recent
surveys at different frequencies to search for new blazars.
We cross-matched the 1$^{st}$ \sw\ XRT Point Source catalog with the
spectroscopic sample of the 9$^{th}$ Data Release of the Sloan Digital
Sky Survey.
Then, we performed further cross-matches with the catalogs
corresponding to the Faint Images of the Radio Sky at Twenty cm survey
and to the AllWISE Data release, focusing on sources with infrared
colors similar to those of confirmed $\gamma$-ray blazars included in
the Second \frm-LAT catalog.
As a result, we obtained a preliminary list of objects with all the
elements needed for a proper blazar classification according to the
prescriptions of the \bz.
We carefully investigated additional properties such as their
morphology and the slope of their spectral energy distribution in the
radio domain, the features shown in their optical spectrum, and the
luminosity in the soft X rays to exclude generic active galactic
nuclei and focus on authentic blazar-like sources.
At the end of our screening we obtained a list of 15 objects with
firmly established blazar properties.
\end{abstract}


\section{Introduction}
\label{sect:01}

Blazars are peculiar radio-loud active galactic nuclei (AGNs) emitting
over the whole electromagnetic spectrum, from the radio band to the
MeV--GeV range of $\gamma$ rays and extending towards the TeV range in
the most extreme cases.
Large-amplitude flux and spectral variability, on short time scales
down to a few hours (\ie,
\citealp{2006A&A...451..435M,2008AJ....135.1384G}), characterize their
emission.
In the radio band they appear as core-dominated, flat spectrum sources.
The discovery of $\gamma$-ray emission from blazars
\citep{1978Natur.275..298S,1999ApJS..123...79H,2011ApJ...743..171A}
established them as a class of extreme particle accelerators which
generate some of the most energetic photons of any known extragalactic
cosmic source.
Within the framework of the AGN Unification Scenario
\citep{1995PASP..107..803U}, these properties are interpreted as non
thermal radiation emitted by blobs of charged particles moving away
from a central supermassive black hole at velocities close to the
speed of light and in directions very close to the line of sight.
In these conditions, the non thermal radiation is relativistically
amplified and becomes dominant with respect to all other components of
thermal origin.
The blazar class includes both BL~Lac objects (BL~Lacs) and Flat
Spectrum Radio Quasars (FSRQs), similar in the shape of their spectral
energy distribution (SED) but different in their optical spectrum.
BL~Lacs display no or very weak emission lines with equivalent width
W$_{\lambda}~\la~5$~\AA\ (\citealp{1991ApJ...374..431S,1991ApJS...76..813S});
conversely, FSRQs reveal evidence of broad emission lines.
According to the Unification Scenario, BL~Lacs and FSRQs are thought
to be the beamed counterparts of low- (FRI) and high-luminosity (FRII)
radio galaxies \citep{1974MNRAS.167P..31F}, respectively.

In recent years the number of all-sky surveys in different energy
bands has considerably increased.
This large amount of data is very helpful for blazar research but
needs to be properly managed.
At this purpose a few, well-defined and stringent criteria for blazar
classification have been defined leading to the
\bz\ \citep{2009yCat..34950691M,2009A&A...495..691M,2010arXiv1006.0922M,2011bzc3.book.....M}.
These criteria are: 1) detection in the radio band down to mJy flux
densities at 1.4~GHz; 2) for FSRQs, a radio spectral index $\alpha_r$
lower than $\alpha_r^{\star}=0.5$ (F($\nu$) $\propto \nu^{-\alpha}$)
measured between 1.4~GHz and 5~GHz: this condition is not required for
BL~Lacs, although most of them have flat spectra; 3) compact radio
morphology or, when extended, with a dominant core and a one-sided
jet; 4) optical identification and analysis of the optical spectrum to
establish the type of blazar; 5) isotropic X-ray luminosity
$L_X~\geq~L_X^{\star} = 10^{43}$~erg~s$^{-1}$ from a point-like
source.
The 4$^{th}$ version of the \bz, available
online\footnote{http://www.asdc.asi.it/bzcat4/}, included 3,149
objects: 1,221 BL~Lacs (BZBs in the catalog, $\sim$38.8\%),
1,707~FSRQs (BZQs, $\sim$54.2\%), and 221 blazars with uncertain
classification (BZUs, $\sim$7.0\%).
270 sources were included among the BL~Lacs despite the lack of an
optical spectrum and were therefore marked as {\it candidates}.
Recent optical spectroscopic campaigns (\eg,
\citealp{2013AJ....145..114L,2014AJ....147..112P,2014AJ....148...66M})
have been carried out to confirm the BL~Lac classification of these
sources; additional campaigns are ongoing.

New, complementary approaches to the blazar classification come from
the evaluation of their color indices in the infrared
\citep{2011ApJ...740L..48M} as well as in the optical-UV bands
\citep{2012MNRAS.422.2322M}.
The discovery of peculiar infrared colors of $\gamma$-ray emitting
blazars \citep{2012ApJ...748...68D} led to the parametrization of the
region that they occupy in the color space of the Wide-field Infrared
Survey Explorer (\wse, \citealp{2010AJ....140.1868W}), named the
\wse\ blazar {\it locus} \citep{2012ApJ...750..138M}.
This parametrization was used to develop a method
\citep{2013ApJS..206...12D} for the association of unidentified
$\gamma$-ray sources detected by \frm-LAT \citep{2009ApJ...697.1071A}
and included in the Second \frm-LAT catalog (2FGL,
\citealp{2012ApJS..199...31N}).
According to this method, the position of a generic IR source with
respect to the {\it locus} can be considered a useful indicator of its
blazar nature.
Recently, the parametrization has been slightly modified to take into
account the results of the AllWISE Data
Release\footnote{http://wise2.ipac.caltech.edu/docs/release/allwise/expsup}
leading to the compilation of an all-sky catalog of candidate
$\gamma$-ray blazars \citep{2014ApJS..215...14D}.
This catalog includes the \wse\ blazar-like radio-loud sources
(WIBRaLS) detected in all the four \wse\ filters.

In this work we provide a list of sources with firmly established
blazar properties after a critical review of a preliminary list of
objects obtained from the cross-match of recent multi-frequency
surveys.
The paper is organised as follows: the guidelines adopted in the
cross-match of the survey catalogs used in this work are illustrated
in Section~\ref{sect:02}.
The validation of a sample of blazars among a preliminary list of
candidates is presented in Section~\ref{sect:03}.
Finally, a discussion of our results is reported in
Section~\ref{sect:04}.

The \wse\ magnitudes in the [3.4], [4.6], [12], and [22] $\mu$m
nominal filters are expressed in the Vega system; the magnitude values
of three \wse\ filters (namely those corresponding to [3.4], [4.6],
and [12] $\mu$m) and of the colors [3.4]-[4.6], [4.6]-[12], and
[12]-[22] derived using these magnitudes, are corrected for Galactic
extintion according to the extinction law presented by
\citet{2003ARA&A..41..241D}.
Unless stated otherwise, we used c.g.s. units throughout and assumed a
$\Lambda$ cold dark matter ($\Lambda$CDM) cosmology with
H$_0$~=~72~km~s$^{-1}$ Mpc$^{-1}$, $\Omega_M$~=~0.26, and
$\Omega_{\Lambda}$~=~0.74 \citep{2009ApJS..180..306D}.

\begin{table*}
\scriptsize
\caption{The list of validated blazars}
\label{table:1}
\begin{center}
\begin{tabular}{ccccccccclccc}
\tableline
~\\                                                                                                                        
name &
$d_{OX}$\tablenotemark{a} &
$R_X$\tablenotemark{b} &  
$d_{OR}$\tablenotemark{c} & 
$d_{OI}$\tablenotemark{d} & 
$S_{1.4}$\tablenotemark{e} & 
$S_{5}$\tablenotemark{f} & 
$\alpha_r$\tablenotemark{g} &   
$r$ &
$z$ &
$F_X$\tablenotemark{h} & 
$L_X$\tablenotemark{i} & 
classification \\
~\\                                                                                                                        
\tableline
~\\                                                                                                                        
J003931.58$-$111102.4  &   1.2    &     4.2   &   0.41  &   0.24   &   296     &  237      &  $+$0.18  &  19.0      &    0.553   &  6.30   &  75.1  & FSRQ   \\
J082438.99$+$405707.7  &   2.1    &     4.7   &   0.51  &   0.09   &   175     &  138      &  $+$0.19  &  17.0      &    0.612   &  0.10   &   1.5  & FSRQ   \\
J082753.69$+$521758.3  &   2.6    &     3.8   &   0.68  &   0.14   &   181     &  292      &  $-$0.38  &  18.9      &    0.338   &  2.50   &   9.2  & FSRQ   \\
J095906.96$+$050958.9  &   1.5    &     4.4   &   0.42  &   0.09   &    47     &   73      &  $-$0.35  &  18.0      &    0.996   &  0.38   &  19.5  & FSRQ   \\
J100033.84$+$132410.8  &   1.7    &     3.9   &   0.20  &   0.13   &    53     &   34      &  $+$0.36  &  16.7      &    1.355   &  0.74   &  82.2  & FSRQ   \\
J103045.22$+$255522.1  &   3.1    &     5.1   &   0.18  &   0.04   &    49     &   50      &  $-$0.02  &  17.1      &    0.692   &  0.24   &   5.0  & FSRQ   \\
J104031.62$+$061721.7  &   1.1    &     3.9   &   0.12  &   0.13   &    39     &   49      &  $-$0.18  &  19.9      &    0.743?  &  0.66   &  15.9  & BL~Lac \\
J110838.98$+$255613.2  &   0.3    &     3.8   &   0.20  &   0.31   &    68     &   73      &  $-$0.06  &  17.5      &    0.732   &  0.80   &  19.0  & FSRQ   \\
J133631.44$+$031423.5  &   5.4    &     4.3   &   0.32  &   0.75   &    62     &  111      &  $-$0.47  &  18.9      &    1.303   &  0.12   &  12.1  & FSRQ   \\
J141238.66$+$484447.1  &   1.6    &     3.8   &   0.16  &   0.13   &    58     &   74      &  $-$0.20  &  18.6      &    0.906   &  0.39   &  15.8  & FSRQ   \\
J142114.05$+$282452.8  &   3.4    &     4.1   &   0.68  &   0.03   &    49     &   40      &  $+$0.16  &  17.8      &    0.538   &  0.46   &   5.1  & FSRQ   \\
J153458.41$+$575625.6  &   2.5    &     4.6   &   0.08  &   0.29   &     7     &   17      &  $-$0.71  &  19.0      &    1.129   &  0.17   &  11.9  & FSRQ   \\
J161541.21$+$471111.7  &   2.9    &     4.6   &   0.23  &   0.09   &    98     &  175      &  $-$0.46  &  17.2      &    0.199   &  1.28   &   1.4  & BL Lac \\
J162805.20$+$252636.8  &   1.8    &     4.7   &   0.14  &   0.08   &   100     &   59      &  $+$0.42  &  18.4      &    0.995   &  0.31   &  15.9  & FSRQ   \\
J170634.12$+$361508.0  &   4.9    &     4.4   &   0.37  &   0.14   &    19     &   27      &  $-$0.28  &  17.9      &    0.917   &  1.06   &  44.2  & FSRQ   \\
~\\                                                                                                                        
\tableline  
\end{tabular}
\end{center}
\tablenotetext{a}{Angular separation between the SDSS and the 1SXPS sources in arcseconds}
\tablenotetext{b}{90\% error radius of the 1SXPS source in arcseconds}
\tablenotetext{c}{Angular separation between the SDSS source and the corresponding FIRST associated counterpart in arcseconds}
\tablenotetext{d}{Angular separation between the SDSS source and the corresponding \wse\ associated counterpart in arcseconds}
\tablenotetext{e}{Radio flux density at 1.4~GHz in mJy}
\tablenotetext{f}{Radio flux density at 5~GHz in mJy}
\tablenotetext{g}{Radio spectral index $\alpha_r$}
\tablenotetext{h}{Unabsorbed X-ray flux (0.3--10~keV) in 10$^{-12}$ erg~cm$^{-2}$~s$^{-1}$}
\tablenotetext{i}{X-ray luminosity (0.3--10~keV) in 10$^{44}$ erg~s$^{-1}$}
\end{table*}

\section{Cross-match of the used survey catalogs}
\label{sect:02}

We started our analysis taking into account, among the criteria for
blazar's classification \citep{2009A&A...495..691M}, those related to
the point-like X-ray emission and to the possibility of analysing the
optical spectrum.
Therefore, we used
TOPCAT\footnote{\underline{http://www.star.bris.ac.uk/$\sim$mbt/topcat/}}
\citep{2005ASPC..347...29T} to cross-match the 1$^{st}$ \sw\ XRT Point
Source catalog (1SXPS, \citealp{2014ApJS..210....8E}) with the
spectroscopic sample of the 9$^{th}$ Data Release of the Sloan Digital
Sky Survey (SDSS-DR9, \citealp{2012ApJS..203...21A}).
The 1SXPS catalog resulted from the analysis of data collected by the
\sw\ X-Ray Telescope (XRT, \citealp{2005SSRv..120..165B}) across 8
years of operations since its launch, up to October 2012.
It included 151,524 soft X-ray point-like sources detected in the
0.3--10 keV band with an approximately uniform sky distribution.
\citet{2014ApJS..210....8E} carried out several checks to filter out
spurious or extended sources and to express the reliability of each
detected source.
As a result, a quality flag which can assume three integer values (0,
1, and 2) to indicate a good, reasonable, or poor detection,
respectively, was assigned to each source.
In our cross-match we considered sources from a cleaned version of the
1SXPS including only detections flagged as good or reasonable, made of
98,762 objects.
We spatially cross-matched this list with the SDSS-DR9 survey catalog
searching for optical counterparts, with available spectra, within the
error circle (at 90\% confidence) of each X-ray source.
At this purpose, we added to this {90\%} error radius a positional
uncertainty of 2\as\ to take into account the typical seeing of the
SDSS.
The list of 5012 SDSS sources that we obtained was then cross-matched
with the WIBRaLS catalog \citep{2014ApJS..215...14D}.
We used an overall matching radius of 3.3\as\ as determined by
\cite{2013ApJS..206...12D} in their search for the optimal value of
the spatial association between blazars in the \bz\ and \wse\ sources,
and obtained 214 objects.
All the WIBRaLS have a radio counterpart.
However, we kept for further investigation only objects with a radio
counterpart listed in the Faint Images of the Radio sky at Twenty~cm
Survey (FIRST, \citealp{1995ApJ...450..559B,1997ApJ...475..479W})
since the angular resolution of this survey allows a better evaluation
of the radio morphology.
In the cross-match between our list of 214 SDSS sources and the latest
version of the FIRST catalog \citep{2012yCat.8090....0B} we adopted a
matching radius of 4\as\ to take into account the positional
uncertainty on the FIRST source (2\as) in addition to the SDSS one.
We obtained 201 objects including 152 blazars already present in the
4$^{th}$ version of the \bz: consequently, we achieved a preliminary
list of 49 blazar candidates whose properties were investigated in
greater detail.

\section{Validation of blazar candidates}
\label{sect:03}

The two main aspects that we took into account in the examination of
the radio emission properties of our candidates were the morphology
and the slope of their SED.
Blazars are characterized by a core-dominated radio morphology; as an
alternative, the additional presence of a one-sided jet can be
accepted.
The evidence of extended structures, and in particular those shaped in
the form of two lobes emerging from the central core following an FRI
or an FRII morphology \citep{1974MNRAS.167P..31F} is the clear
footprint of a radio galaxy instead of a blazar.
Other irregular extended structures, adequately resolved by the FIRST
survey, were not accepted either.
As a result, we excluded from our preliminary list 11 radio galaxies,
mainly characterized by a FRII morphology.

Then we estimated the slope in the SED at radio frequencies of
the remaining 38 objects finding a steep spectrum for 13 of them.
In our list we also found 10 objects with no measurements of flux
density $S$ at radio frequencies different from 1.4~GHz; moreover, the
corresponding $S_{1.4}$ values are quite low, at the level of a few
mJy.
The optical spectra of all these 10 sources show evident and broad
emission lines typical of quasars, for which a flat radio spectrum is
required (see Section~\ref{sect:01}).
In these conditions the impossibility of measuring the slope of the
radio spectrum prevented them from being safely classified as blazars,
despite any other requirement was fulfilled.

Blazars can be distinguished from other kind of sources through the
comparison of the continuum emission from the active nucleus with
respect to the host galaxy contribution.
A spectroscopic analysis in the optical band is able to show the
mutual strength of these two components.
At this purpose the Ca~H\&K break absorption
\citep{1991ApJS...76..813S,1996MNRAS.281..425M}, if present, is one of
the most frequently investigated features.
A weak non-thermal continuum emission, reflecting in a very sharp Ca~H\&K
break, is the footprint of a weak central engine and probably shares
little with the blazar nature, characterized by much more extreme
properties.
Recently, \citet{2012MNRAS.422.2322M} have pointed out that, for
objects with $z~\leq$~0.5, the amount of the Ca~H\&K break absorption
in the SDSS spectra can be determined by computing their $(u-r)$ color
index from the magnitudes of their SDSS photometry adopting the
formula:
\begin{equation}
\label{ur}
(u-r) = (u-r)_{obs} - 0.81 A_r
\end{equation}
where $A_r$ is the extinction in the $r$~band.
In their analysis, dominant nuclear activity in extragalactic sources
has been found to correspond to $(u-r)$ values lower than a threshold
$(u-r)^{\star}$~=~1.4.
Among the 15 remaining objects only two are at redshift $z~\leq$~0.5,
and their $(u-r)$ color index is lower than $(u-r)^{\star}$.

\begin{figure}[htpb]
\begin{center}
\includegraphics[width=65mm, angle=-90]{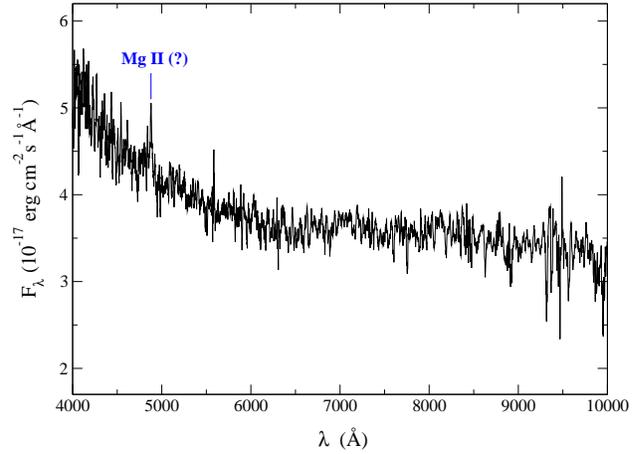}
\end{center}
\caption{The SDSS-DR9 optical spectrum of SDSS~J104031.62$+$061721.7.}
\label{fig:01}
\end{figure}

The list of 15 firmly established blazars selected at the end of our
investigation is shown in Table~\ref{table:1}.
Each source has been addressed using its SDSS name (column~1).
The angular separation $d_{OX}$ between the SDSS and 1SXPS coordinates
is reported (column~2) together with the 90\% error radius $R_X$ of
the 1SXPS source (column~3) provided by \cite{2014ApJS..210....8E}.
We have also reported the angular separation of the SDSS source from
the FIRST ($d_{OR}$, column~4) and the \wse\ ($d_{OI}$, column~5)
associated counterparts, respectively.
The values of the radio flux density $S_{1.4}$ (column~6) have
been taken from the most recent version (2012, February 16) of the
FIRST survey catalog.
The values $S_{5}$ (column~7) come from the Green Bank 6~cm Radio
Source Catalog (GB6, \citealp{1996ApJS..103..427G}) with the only
exception of SDSS~J003931.58$-$111102.4 for which the value found in
the Parkes-MIT-NRAO survey (PMN, \citealp{1994ApJS...91..111W}) was
reported.
The values of the radio spectral index $\alpha_r$ between these two
frequencies (column~8) have been calculated using the formula:
\begin{equation}
\label{alfaradio}
\alpha_r = -1.81 \cdot Log~(S_{5}/S_{1.4}).
\end{equation}
The values of both the magnitude in the $r$~filter (column~9) and the
redshift $z$ (column~10) come from the SDSS-DR9.
One of the redshift values, corresponding to
SDSS~J104031.62$+$061721.7, is reported with a question mark as we
found a flag in the optical spectrum.
The flag is due to the fact that the $z$ value provided in the SDSS is
based on a single, faint emission line.
Assuming that this feature in the optical spectrum (Fig.~\ref{fig:01})
can be attributed to Mg~II we carried out a further check on the
redshift measurement.
The value that we computed, which is reported in Table~\ref{table:1},
is very similar to the one ($z$=0.735) given in the SDSS.
The values of the unabsorbed flux $F_X$ over the 0.3--10~keV band
(column~11) come from \citet{2014ApJS..210....8E}.
The corresponding values of the X-ray luminosity $L_X$ (column~12)
were computed from the formula $L_X~=~4~\pi~(d_L)^2~\cdot~F_X$, where
the luminosity distance $d_L$ was derived from the redshift $z$ using
the web
calculator\footnote{\underline{http://www.astro.ucla.edu/$\sim$wright/CosmoCalc.html}}
by \cite{2006PASP..118.1711W}.
The values that we obtained are well above the threshold $L_X^{\star}$
established in the \bz\ (Section~\ref{sect:01}), by at least an order
of magnitude.
The quality flag of all the corresponding 1SXPS sources is equal to 0
meaning a good quality X-ray detection (Section~\ref{sect:02}).
Finally, in column~13 we reported the proposed classification as a
BL~Lac or as a FSRQ, mainly driven by the features shown in their
optical spectrum.

\begin{figure}[thpb]
\begin{center}
\includegraphics[width=65mm,angle=-90]{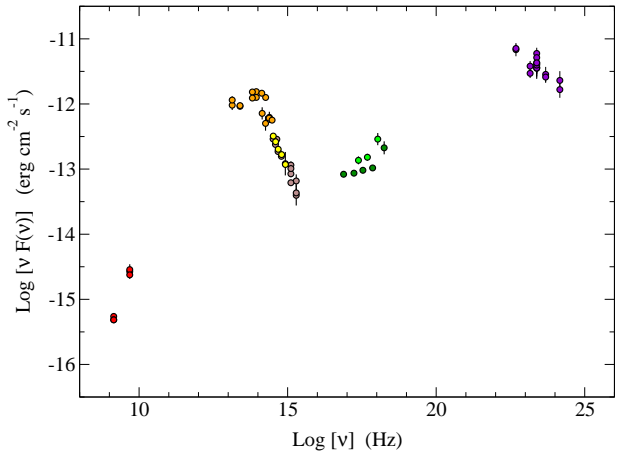}
\end{center}
\caption{The SED of SDSS~J104031.62$+$061721.7. Red circles correspond
  to radio data at 1.4~GHz and 5~GHz; orange circles correspond to
  infrared data from \wse, 2MASS (\citealp{2006AJ....131.1163S}) and
  UKIDSS (\citealp{2007MNRAS.379.1599L}); SDSS data (yellow) are
  superposed to various data from the NED database (brown); X-ray data
  come from the \sw~(light green) and the {\it XMM}-Newton (green)
  satellites; purple circles correspond to \frm~data.}
\label{fig:02}
\end{figure}

\section{Discussion}
\label{sect:04}

Blazars are certainly not the only sources characterized by a
multifrequency emission.
In recent years their peculiar properties have been well defined by
restrictive criteria: the need for establishing them has become quite
urgent as a consequence of the rapid increase in the lists of
candidates, a large number of which is simply the result of automatic
cross-matches of the various all-sky surveys obtained by spacecraft
telescopes at different energy bands.
For this reason, a careful check of the overall properties is highly
recommended to support any candidate before validating the blazar
classification.
At the end of our screening procedure, that includes some steps that
could not be carried out automatically, we obtained a list of 15
blazars.
We verified that the SED of all our blazars is consistent with the
characteristic double-bumped shape, the former attributed to
synchrotron radiation and the latter to inverse Compton emission.
One of them is reported as example in Fig.~\ref{fig:02}. 
All the values of the radio spectral index $\alpha_r$ are strictly
consistent with a flat radio spectrum.
According to the classification that we propose all the sources
  of our list, with two exceptions, are FSRQs with a mean redshift
value $\bar{z}$ = 0.851.
FSRQs are generally found among low synchrotron peak (LSP,
\citealp{2010ApJ...716...30A}) sources: in these conditions, the
detection in the soft X rays corresponds to the inverse Compton
emission.

\begin{figure}[hbpt]
\begin{center}
\hspace{-24mm}
\includegraphics[width=64mm]{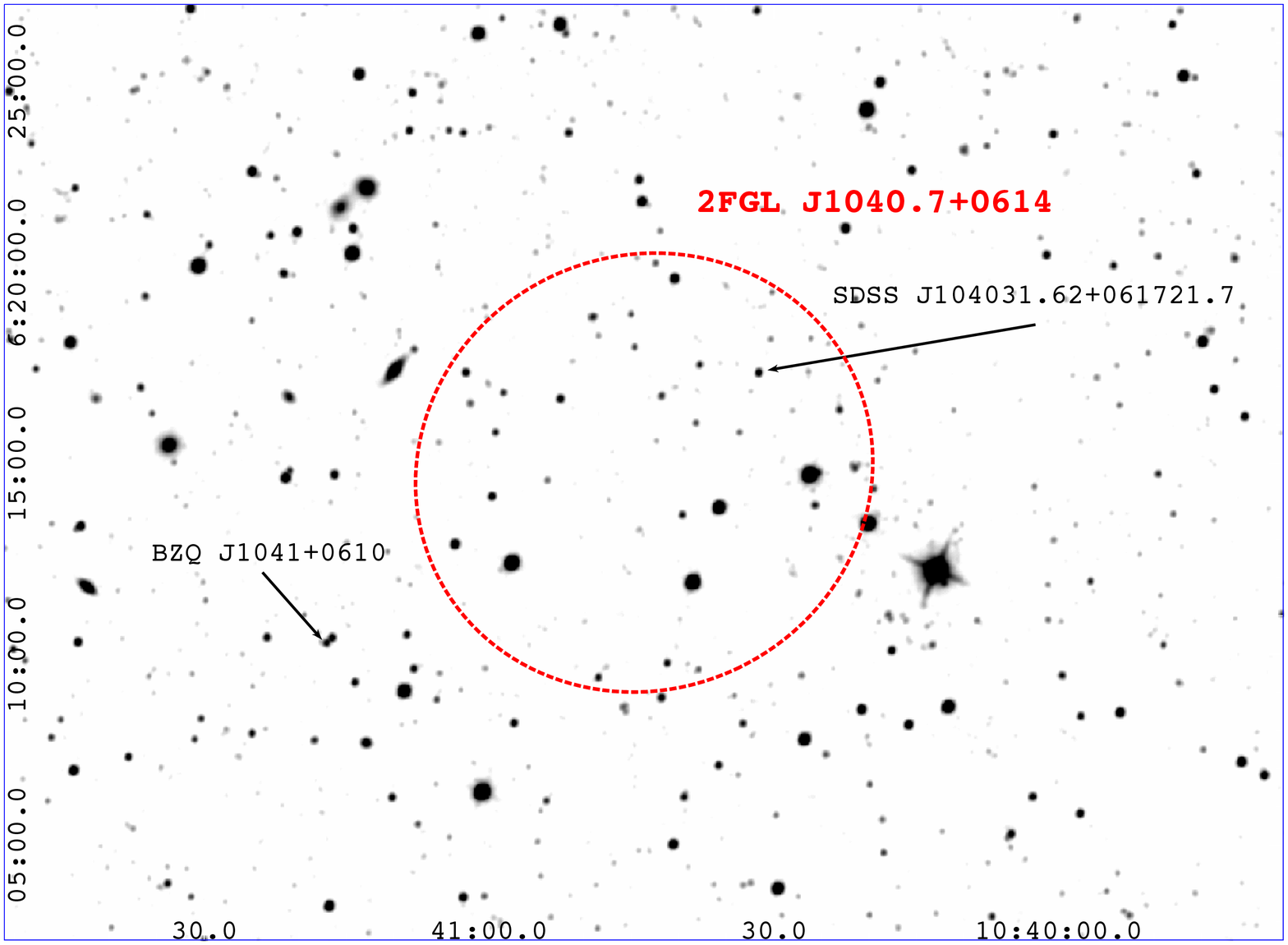}
\end{center}
\caption{\wse\ 3.4~$\mu$m sky map centered at the position of
  {\footnotesize 2FGL~J1040.7$+$0614}. The BL~Lac object found by our
  procedure, {\footnotesize SDSS~J104031.62$+$061721.7}, is within the
  error region of the $\gamma$-ray source (red crossed line) at
  variance with {\footnotesize BZQ~J1041$+$0610}, the counterpart
  associated in the 2LAC.}
\label{fig:03}
\end{figure}

In the optical spectrum of {\footnotesize SDSS~J104031.62$+$061721.7}
(Fig.~\ref{fig:01}) there is evidence of continuum emission with no
broad emission lines, suggesting a BL~Lac classification.
Considerable variability in the optical band was reported by
\cite{2009ApJ...705...46B}.
We note (Fig.~\ref{fig:03}) that this source is within the 95\% error
circle $\theta_{95} = 5.88\arcmin$ of the $\gamma$-ray source
{\footnotesize 2FGL~J1040.7$+$0614}; the angular separation between
the SDSS and the 2FGL coordinates is 3.96\arcmin.
Conversely, the counterpart 4C~$+$06.41 (named {\footnotesize
  BZQ~J1041$+$0610} in the \bz) currently associated to this
$\gamma$-ray source in the Second LAT AGN Catalog (2LAC,
\citealp{2011ApJ...743..171A}) is outside of the 2FGL error circle, at
a distance of 9.45\arcmin; for this reason, their association is given
with a low bayesian probability.
Therefore, we consider the eventuality that a contribution to the
$\gamma$-ray emission corresponding to {\footnotesize
  2FGL~J1040.7$+$0614} might be attributed to {\footnotesize
  SDSS~J104031.62$+$061721.7}, possibly in addition to {\footnotesize
  BZQ~J1041$+$0610}.

The other source that does not show broad emission lines in its
optical spectrum is {\footnotesize SDSS~J161541.21$+$471111.7}.
The value that we found for its color index $(u-r)$~=~1.34, even if
lower than the threshold $(u-r)^{\star}$ established by
\cite{2012MNRAS.422.2322M}, suggests a relatively weak non-thermal
contribution with respect to the host galaxy.
Moreover, this source was recently addressed by
\citet{2013ApJS..206...13M} as a possible counterpart of the
unidentified $\gamma$-ray source (UGS) {\footnotesize
  2FGL~J1614.8+4703} detected by \frm-LAT.

If we consider the totality of blazars selected by our method,
including the 152 already reported in the 4$^{th}$ version of the \bz,
in our list of 167 objects we found 72~BZBs ($\sim$43.1\%), 85~BZQs
($\sim$50.9\%) and 10~BZUs ($\sim$6.0\%), respectively.
These rates are in agreement with those corresponding to the whole
blazar population as reported in the \bz\ (Section~\ref{sect:01}).

\acknowledgments

The work is supported by the ASI grant I/004/11/0 and by the NASA
grants {\scriptsize NNX12AO97G} and {\scriptsize
  NNX13AP20G}. R. D’Abrusco acknowledges the SI competitive research
grant ``A New Astro-Archeology Probe of the Merging Evolution of
Galaxies'' for support. R. D’Abrusco's work was also partially
supported by the Chandra X-ray Center (CXC), which is operated by the
Smithsonian Astrophysical Observatory (SAO) under NASA contract
NAS8-03060. The work by G. Tosti is supported by the ASI/INAF contract
I/005/12/0. TOPCAT and SAOImage DS9 were used extensively in
this work for the preparation and manipulation of the tabular data and
the images.  This research has made use of data obtained from the High
Energy Astrophysics Science Archive Research Center (HEASARC) provided
by NASA's Goddard Space Flight Center; the SIMBAD database operated at
CDS, Strasbourg, France; the NASA/IPAC Extragalactic Database (NED)
operated by the Jet Propulsion Laboratory, California Institute of
Technology, under contract with the National Aeronautics and Space
Administration. Part of this work is based on archival data, software
or on-line services provided by the ASI Science Data Center. This
publication makes use of data products from the Wide-field Infrared
Survey Explorer, which is a joint project of the University of
California, Los Angeles, and the Jet Propulsion Laboratory/California
Institute of Technology, funded by the National Aeronautics and Space
Administration. This work made use of data supplied by the UK Swift
Science Data Centre at the University of Leicester. Funding for
SDSS-III has been provided by the Alfred P. Sloan Foundation, the
Participating Institutions, the National Science Foundation, and the
U.S. Department of Energy Office of Science. The SDSS-III web site is
http://www.sdss3.org/.  SDSS-III is managed by the Astrophysical
Research Consortium for the Participating Institutions of the SDSS-III
Collaboration including the University of Arizona, the Brazilian
Participation Group, Brookhaven National Laboratory, University of
Cambridge, Carnegie Mellon University, University of Florida, the
French Participation Group, the German Participation Group, Harvard
University, the Instituto de Astrofisica de Canarias, the Michigan
State/Notre Dame/JINA Participation Group, Johns Hopkins University,
Lawrence Berkeley National Laboratory, Max Planck Institute for
Astrophysics, Max Planck Institute for Extraterrestrial Physics, New
Mexico State University, New York University, Ohio State University,
Pennsylvania State University, University of Portsmouth, Princeton
University, the Spanish Participation Group, University of Tokyo,
University of Utah, Vanderbilt University, University of Virginia,
University of Washington, and Yale University.

%
\bibliographystyle{spr-mp-nameyear-cnd}  
\bibliography{maselli}        

\end{document}